\documentstyle[preprint,aps,prbbib]{revtex}
\begin{document}
\title{Influence of the exchange reaction on the
electronic structure of GaN/Al junctions}
\author{S. Picozzi and A. Continenza}
\address{Istituto Nazionale di Fisica della Materia (INFM) -
Dipartimento di Fisica \\
Universit\`a degli Studi di L'Aquila, 67010 Coppito (L'Aquila), Italy }
\author{S. Massidda}
\address{Istituto Nazionale di Fisica della Materia (INFM)
 - Dipartimento di Scienze Fisiche \\ Universit\`a degli
Studi di Cagliari - 09124 Cagliari, Italy \\}
\author{A. J. Freeman}
\address{Department of Physics and Astronomy and Materials Research Center\\
Northwestern University, Evanston, IL 60208 (U.S.A.)\\ and}
\author{N. Newman}
\address{Department of Electrical and Computer Engineering\\
Northwestern University, Evanston, IL 60208 (U.S.A.)\\}

\maketitle
\begin{abstract}
Ab-initio full-potential linearized augmented plane wave
(FLAPW) calculations have been used to study the influence
of the interface morphology and, notably, of the
exchange reaction on the electronic properties of Al/GaN
(100) interfaces.
Although the detailed mechanism
is not understood, the exchange reaction has been purported
to influence the Schottky barrier height as a result of the
formation of an interfacial Ga$_{x}$Al$_{1-x}$N layer.
In particular, the effects
of interface structure  ({\em i.e.}  interfacial bond lengths,
semiconductor surface polarity,
and reacted intralayers) on the SBH at the Al/GaN
(001) junction are specifically addressed.   Thus, the electronic
structure of the
following atomic configurations have been investigated
theoretically: {\em (i)} an abrupt, relaxed GaN/Al interface;
{\em (ii)} an interface which has undergone one monolayer of
exchange reaction; and interfaces with a monolayer-thick
interlayer of {\em (iii)} AlN and {\em (iv)} Ga$_{0.5}$Al$_{0.5}$N.
The exchange reaction is found to be exothermic with an
enthalpy of ~0.1 eV/atom.
We find that the first few layers of semiconductor are
metallic due to the tailing of metal-induced-
gap-states (MIGS); therefore, the presence of a monolayer--thick
 interfacial alloy layer does not result in
an enhanced bandgap near the interface.
Intermixed interfaces are found to pin the interface Fermi level at a
position not significantly different from that of an abrupt
interface.
Our calculations also show that the interface
band line--up is not strongly dependent on the interface
morphology changes studied.  The p-type SBH is
reduced by less than 0.1 eV if the GaN surface is
Ga-terminated compared to the N-terminated one.  Moreover,
we show that both an ultrathin Ga$_x$Al$_{1-x}$N ($x$ = 0, 0.5)
intralayer and a Ga$\leftrightarrow$Al atomic swap at the
interface does not significantly affect the Schottky barrier
height.
\end{abstract}
\leftline{PACS 71, 73, 73.40.Sx, 73.61.-r}

\section{Introduction}

In the last few decades, a significant number
of studies have investigated metal/semiconductor
interfaces, due to the key role played by ohmic
and Schottky contacts
in technological III-V semiconductor devices.  However,
from both the experimental and the theoretical points of
view, the fundamental mechanism involved in the Schottky
barrier formation has not been fully elucidated
\cite{monchrev,Rhodrev,Brillrev,ohmic}.

Al / III-V semiconductor junctions have been the most
extensively studied interfaces to date.  This has occurred due
to both practical and fundamental reasons. Al contacts are
inexpensive to manufacture and have stable electrical,
chemical and structural characteristics at moderate
temperatures\cite{Ghandi}.  Al contacts also are of practical interest
since they are commonly used in III-V commercial device
technology, particularly as the gate of field effect devices.

In the field of ab-initio computational physics, many
works have addressed band alignment at semiconductor
heterojunctions\cite{het}, whereas few calculations have
studied the barrier height of metal/semiconductor interfaces.
Al has been used almost exclusively as the metal layer in
theoretical investigations of metal/semiconductor contacts
\cite{Louie,Zunger,Ihm,Swarts,Chadi,Charlesworth}.
This arose because
 Al is found to be advantageous since it
lacks d-electrons,
forms a free-electron metal,
and lattice matches a number of III-V semiconductors.
Recently, the GW method has been used to study the Al/
GaAs interface\cite{Charlesworth}.

More recently, theoretical work on III-V
interfaces demonstrated that the pressure-induced barrier
height changes can be used to critically determine the nature
of the states which pin metal/III-V interfaces
\cite{baldereschini,vanP}.  Experimental
measurements of the pressure-induced change in Al/GaAs
and Al/AlGaAs barrier heights are consistent with defect-
free interfaces\cite{baldereschini,Phatak},
validating Al/III-V interfaces as a model system.  In
contrast, Au, a metal which reacts with GaAs
to release near-interfacial As at GaAs interfaces, exhibits
pressure-induced barrier height changes consistent with
an interface decorated with a deep-level point-defect, such as
the As antisite (As$_{Ga}$) \cite{Phatak}.
Because GaN solid is significantly more stable than
GaAs, the Fermi-level pinning position of Al/GaN interfaces
would also be expected to be determined by the properties
of interfaces without deep-level defects.

In the study reported here, we start an investigation of an important
aspect of Schottky barriers  that only recently has begun to be
explored using modern electronic structure calculations:
 the role
that  interface chemical exchange plays in the electrical
and electronic properties of interfaces.
Al/GaN interfaces were used in this study due to the well
known chemistry and atomic structure of the interface.  GaN
\cite{bermudez}, GaAs\cite{Skeath},
and InP \cite{Kende}, among others, are found to exhibit an exchange
reaction in which Al replaces the Group III element at the
surface.  The extent of the reaction is found to be of the order
of a monolayer for unannealed contacts and increases at
elevated annealing temperatures\cite{Chambers}.  Changes in
the electrical properties, as reported in a large number of
 III-V studies, have
attributed the increase in the barrier for n-type
semiconductors to the increased bandgap\cite{Sun}.  However,
a more recent study on both n-type and p-type contacts indicates that
the observed behavior can be attributed to a shift in interfacial
Fermi level pinning position, rather than due to the formation
of an increase in interfacial bandgap\cite{NewmanAl}.  A strong
fundamental
understanding of the influence of the chemical reaction on the
interface electronic structure has not yet been established.

Density functional calculations are able to
quantitatively address many of the questions including the
energetics of the exchange reaction, the nature and extent of
the MIGS (metal-induced gap states) and the role of the
larger-bandgap AlGaN interface layer on the interface
electronic structure.
In a previous work\cite{silvia5}, we performed {\em ab-initio}
calculations to determine the electronic properties of the
[001] ordered atomically abrupt N-terminated
XN/Al (X~=~Ga, Al) interface,  focusing mainly on the
Schottky Barrier Height and the resulting interface states.
In the work reported here, the structural and electronic properties of
several Al/GaN interface configurations are investigated.
In particular, the effects
of interface structure  ({\em i.e.}  interfacial bond lengths,
semiconductor surface polarity,
and reacted intralayers) on the SBH at the Al/GaN
(001) junction are specifically addressed.
It should be noted that some experimental results \cite{bermudez}
are available for the (0001) wurtzite GaN/Al interface and that some caution
has to be taken in comparing our results with the experiments, since the
polarization effects and the different coordinations of the surface bonds may
both play a role in determining the structural and electronic properties.

Starting from the atomically abrupt N-terminated
GaN/Al junction (in the following
denoted as the A structure), we consider four other different interface
morphologies:
{\em (i)} the ideal, abrupt, relaxed Ga-
terminated GaN/Al interface (B); {\em (ii)} the  configuration  with an AlN
intralayer (C); {\em (iii)} the  configuration in which an Al atom has
undergone an exchange reaction with a surface Ga atom (D) and
finally  {\em (iv)} the  configuration with a
Ga$_{0.5}$Al$_{0.5}$N alloy layer at the interface (E).  Our calculations
show that the interface band line--up is not strongly
dependent on the interface morphology.  In particular, we find
that the exchange reaction is found to be exothermic with an
enthalpy of ~0.1 eV/atom.   We show that both an ultrathin
Ga$_x$Al$_{1-x}$N (x= 0, 0.5) intralayer and a
Ga$\leftrightarrow$Al atomic swap at the interface does not
significantly affect  the $p$--type SBH.
The p-type SBH is smaller by 0.1 eV for a Ga-terminated surface
than a N-terminated one.

In Sec. \ref{techn}, we describe the theoretical method
and the atomic configurations used in this study. In Sec. \ref{results},
we discuss our
results, focusing in Subsec. \ref{struct} on the
structural properties, ({\em i.e.} equilibrium atomic distances,
nature of the interface bonds -- metallic versus covalent)
 whereas in Subsec. \ref{elec} we discuss
effects of the interface morphology on the interface states
and the resulting SBH.
Finally, our results are summarized in Sec. \ref{concl}.

\section{Computational method and atomic  configurations used in
calculations}
\label{techn}
First principles full-potential linearized augmented plane
wave (FLAPW \cite{flapw}) calculations within the local density
approximation to density functional theory were used in this
work\cite{HK64}.  Plane waves with wave vector up to
$K_{max}$ = 4.6 a.u., leading to about 5000 basis functions,
were used.  Angular momenta up to $l_{max}$ = 8 in the
muffin tin spheres ($R_{\rm Ga}$ =$R_{\rm Al}$ = 2.1 a.u.
$R_{\rm N}$ = 1.4 a.u.) for both wavefunctions and charge density
were used in the self-consistency cycles.  Three special k-points,
obtained following the Monkhorst-Pack scheme \cite{MP}, were used to
sample  the tetragonal Brillouin zone.

The widely adopted supercell approach was used to simulate
the different types of interfaces.  In particular, we performed
{\em ab-initio} calculations for supercells containing
15 GaN + 7 Al  (13 GaN + 7 Al) layers in the N-terminated
(Ga-terminated) case.  The thickness of these layers
is sufficiently large that the metal and semiconductor
layers farthest from the interfaces have properties similar
to the bulk (see below).

The method employed to evaluate the SBH uses the metal and
semiconductor atomic core levels as reference energies located
sufficiently far from the interface to have bulk-like bandstructure.
In particular, the SBH can be expressed as
$\Phi_{B_p} = \Delta b + \Delta E_b$, where $\Delta b$ and
$\Delta E_b$  indicate an ``interface" and ``bulk" contribution, respectively
(see Ref. \onlinecite{silvia5} for details in the GaN/Al case).
This simple procedure is commonly used in {\em ab-initio} all-electron
calculations to evaluate the interface band line-up at semiconductor
heterojunctions or metal/semiconductor junctions.
We considered all the structures for a cubic
({\em i.e.} zincblende) [001] ordered
GaN substrate, with a calculated lattice constant
$a_{subs}$ = $a_{GaN}$ = 8.47 a.u. \cite{nota1}.
Further structural details can be found in Ref.
\onlinecite{silvia5}. All the structures considered are shown in
Fig. \ref{figs} and summarized in Table \ref{struc}.



The choice of these structures (in particular the C, D and E systems)
has been suggested by
 a recent surface investigation
of Al films deposited on wurtzite GaN(0001)-(1x1) surfaces,
\cite{bermudez} reporting that the interface
is very reactive. It was experimentally observed, in fact,  that
metallic--Ga was released from the interface after
monolayer depositions of Al.  This suggests that an exchange reaction
occurs in which the Al atom replaces the group III element
within the semiconductor, in analogy to that found for GaAs
\cite{Skeath} and InP\cite{Kende}.
After repeated cycles of Al deposition and annealing,
evidence was found \cite{bermudez} for the presence of a
Ga$_x$Al$_{1-x}$N alloy interfacial layer.

indicated by


In Table \ref{struc} (structure E)
the interface plane
X$_i$ indicates  the fcc cationic
sites  are occupied either by  Ga or  Al
(50 $\%$ of Ga and 50 $\%$ of Al),
with an ordered superstructure which aims to simulate the
Ga$_{0.5}$Al$_{0.5}$N alloy. In this case,
we consider a supercell  with doubled dimensions in the $x-y$ plane.
The resulting increase in the computational cost was compensated for by
reducing the
thickness of the supercell ({\em i.e.} 11 GaN + 5 Al layers). This
is expected to affect the final value of the SBH, since
bulk conditions in the thinner
GaN and Al regions may not be completely recovered (see below).
However, the effects of the alloy Ga$_{0.5}$Al$_{0.5}$N intralayer on the
structural and electronic properties can be determined
by comparing the results to the ideal atomically abrupt N-terminated
GaN/Al interface having the same size and in-plane dimensions;
the only difference between them being  the last cationic
plane of the semiconductor side which is made of
50$\%$ Ga and 50$\%$ Al
atoms and all Ga atoms, in the
first and second case,  respectively.

For configurations A-D, structural relaxations along the [001] direction
were performed according to {\em ab-initio} atomic forces \cite{alefor}.
Other degrees of freedom including in-plane atomic relaxations and
formation of extended defects, such as dislocations, were
neglected.  In the case of configuration E, due to onerous computational
costs for the in-plane doubled supercells, structural minimization was not
performed.  Instead, the calculated interplanar distance from structure
A was used.  Despite this limitation, the comparison of the results for
configurations A-E can be used to determine the dependence of the
interfacial electronic properties on the chemistry and structure of the
interface.

\section{Results and discussion}
\label{results}
\subsection{Stability and structural  properties}
\label{struct}

with the
supercells

Since the supercells do not have the same number of each
atomic species, a fully consistent determination of the relative stability
of the configurations is in general not possible from the superlattice
calculations alone \cite{notaform}.  The problem can be avoided by obtaining
the chemical potentials $\mu$ of the atomic species of pure solids
from independent
total energy calculations. However, the problem of biggest
interest in the present contest is the one of the exchange reaction,
and we can get the relevant figure by comparing the ideal N-terminated
interface (A) with the
structure showing the Ga$\leftrightarrow$Al swap (D), since the two supercells
contain the same number of atoms of the same species.
As for the Ga vs N termination, this problem is of more concern in the study of
the GaN surface, and has already been addressed in the
literature\cite{neugebauer}.
As  shown in Ref.\onlinecite{silvia5} and mentioned earlier in this work,
the atomic forces acting on the interface Al$_i$ in  configuration A do not
decrease the  $z_{Al_i-N_i}$  pushing the Al towards the GaN
region.  Instead, we find an equilibrium structure having a bond length
$d_{Al_i-N_i}$ larger than both
$d_{Al-N}^{bulk}$ and  $d_{Ga-N}^{bulk}$
(where $d_{X-N}^{bulk}$ (X = Ga, Al) denotes the bulk bond length
of XN).
However, our calculations show that the total energy of configuration
D is lower (by about 0.1 eV/atom) than the total energy of
structure A, confirming that the Ga$\leftrightarrow$Al exchange
reaction observed experimentally \cite{bermudez} is energetically
favorable. Intuitively, this can be understood by noting that the AlN
bond has a significantly higher bond energy than GaN.
We note that a further displacement of Ga inside the Al overlayer will imply a
more ionic N-Al interface bond (greater cohesive energy), and one
more partially covalent Ga-Al bond on the metal side.
We can therefore reasonably expect that such an exchange
will lead to a further total energy reduction.
The heat of formation for the
Al/GaN exchange reaction is significantly less
than the values predicted for the Al/GaAs system by Ihm (0.48 eV)
\cite{Ihm} and Swarts (0.62 eV)\cite{Swarts}.


The theoretically-determined interplanar distances along the
[001] direction are summurized in Table \ref{struc}.
Since the electronic properties of configuration A have been
discussed in a previous work\cite{silvia5}, we describe these
results only in comparison with the other configurations.
In the atomically abrupt interface B,
the interplanar distance in the last GaN plane
($\Delta z_{Ga_i-N_i}$~=~2.23~a.u.)
is larger than in  bulk GaN (2.12~a.u.), which indicates that the
interface Ga$_i$N$_i$
covalent bond is weakened with respect to the  GaN bond in the bulk.
This can be explained in terms of the
{\em half--metal/half--semiconductor} character of the interface Ga$_i$ atom.
In fact, as already pointed out in Ref.\onlinecite{silvia5} for the
A configuration,
the interface cation ({\em i.e} Al$_i$
 and Ga$_i$ in the A and B case, respectively) forms with the other Al atoms
 a metallic bond, which
reduces its possibility of full
$sp^3$ hybridization, thus weakening the covalent character of the
 N-cation bond.
A similar mechanism is also responsible for making the
interplanar distance $\Delta z_{N_i^2Al_i^2}$ larger for the
C configuration than $\Delta z_{Al_i^1N_i^2}$.
In order to substantiate this, we report in
Fig. \ref{chplot} (a) and (b)
the charge density contour plots relative to the
B configuration for the bulk Al atoms and the interface
Ga$_i$-Al$_i$  atoms, respectively.   The interface
Ga$_i$-Al$_i$ bond is predominantly metallic,  although it is slightly
different from the pure Al-Al bond, due to different Ga and
Al electronegativities (compare Fig. \ref{chplot}
(a) and (b)).

In Fig.\ref{chplot} (c) we  report the charge density contour plots
for the interface Ga$_i^2$
and Al$_i^2$ atoms in configuration D; the similarity of panels (b) and (c)
in Fig. 1 shows that the Ga atom released in the swap mechanism tends to
forms metallic bond with the  deposited Al atoms.
This is consistent with the results from  {\em ab-initio}
atomic forces,  which give an interplanar distance
for  interface D nearly equal (within 1$\%$)
to that in the B configuration (see Table \ref{struc}).
We note that the results obtained in the present work for configuration B
are similar to those obtained from {\em ab-initio}
pseudopotential calculations performed for the
Ga-terminated Al/GaAs(001) interface
\cite{dandrea2,ohno}: the top Ga atomic layer was
found to relax outwards, thus elongating the interface Ga-As bonds, as a
result of metallization of the Ga-Al layer.

Table I shows  that in the fully relaxed C configuration, the
N$_i^1$ - Al$_i^1$ and Al$_i^1$ - N$_i^2$ interplanar distances
($\Delta z_{Al_i^1 - N_i}\:\sim$ 2.12 a.u.) are equal to the
interplanar distance
($\Delta z_{Ga_i - N_i}\:\sim$ 2.12 a.u.) in configuration A.
On the other hand,   the LDA-calculated equilibrium
interplanar   distance
for tetragonal AlN with the in--plane lattice constant
fixed to that of the  GaN substrate is
$\Delta z_{Al-N}^{tetrag}$ =  2.00 a.u.,  leading to a
 bond length  $d^{tetrag}_{Al-N}$ = 3.60 a.u..  \cite{silvia5}

Therefore, our structural results regarding configuration C suggest
that the  AlN intralayer, constrained on a GaN substrate, can not
attain the equilibrium tetragonal bond length
within one layer.  Instead, when
Al replaces the surface Ga atom, the Al is found to
occupy the same position without significant structural relaxation.
Based on these results, we can reasonably expect a similar situation
in the case of configuration E, with the
Al atoms of the  Ga$_{0.5}$Al$_{0.5}$N layer substituting  50 $\%$ of the
Ga atomic sites
without appreciably modifying the bond lengths, thus excluding
the  possibility of {\em buckling} effects.
Hence,  the unrelaxed E configuration, whose
interplanar distances are set to those of configuration A,
should not be
radically different from the fully relaxed structure.

\subsection{Schottky Barrier Heights}
\label{elec}
This subsection focuses on the influence of the interface morphology
on the SBH.
We recall that for the GaN/Al interface we found
that the metal induced gap states (MIGS) tail into the
semiconductor side with a decay length of about
$\lambda \sim$ 3.5 a.u.\cite{silvia5};
in fact, in the present cases we find that the first few
layers of the semiconductor
have metallic character.  For this reason,
the presence of a monolayer-thick  interfacial
alloy layer does not result in an enhanced
bandgap near the interface.  As a consequence, in this case
changes in the electrical properties arise from
shifts in the interface Fermi-level pinning
position within the GaN bandgap.
Table \ref{schbar} lists our theoretical results for the
$p$-type SBH
\mbox{$\Phi_{B_p}\:=\:E_F\:-\: E_{VBM}$}
calculated as  the difference between the Fermi level ($E_F$) and the
GaN valence band maximum ($E_{VBM}$) sufficiently far from the
interface.
Note that the $n$-type SBH, $\Phi_{B_n}$, can be obtained from the
$p$-type SBH through the
following relation
\begin{center}
$\Phi_{B_n}$ = $E_{gap}$ - $\Phi_{B_p}$
\end{center}
 $E_{gap}$ being the experimental semiconductor band gap energy
 ($E_{gap}^{expt}$(GaN) = 3.39 eV \cite{foresi,bermudez}).

We point out that all the results
reported in the first row of Table \ref{schbar} refer to supercells with
15 GaN + 7 Al layers.  In order to check the convergence of
our SBH values as a function of the cell dimensions, we also performed
calculations for a larger supercell, containing 19 GaN + 9 Al layers.
>From this case \cite{silvia5}, we obtained variations of the SBH of
less than 0.02 eV, showing that bulk conditions are well recovered also
in the smaller 15 GaN + 7 Al supercell.  Note also that for  configuration A
we used an even smaller cell size (11 GaN + 5 Al layers - see the second
row in Table \ref{schbar}) in order to compare with configuration E, which,
due to the in-plane doubling of the cell, was calculated with a shorter
periodicity along the growth direction.  The  different value (by about 0.2 eV)
obtained for the 15 GaN + 7 Al (configuration A, first row) with respect to the
11 GaN + 5 Al (configuration A, second row) is  related to
the insufficient thicknesses considered.  However, this is not going to affect
our conclusions, since in this case we are interested in SBH trends rather
than
 absolute values.  The results shown in Table \ref{schbar} do not include  {\em
quasi-particle} corrections
($\Delta_{QP}\:\approx\:$0.1 eV \cite{silvia5})
related to the difference in screening on the two sides of the  interface.
This correction should be added to the $p$-type SBH value before a direct
comparison to experiment is made.

Our calculations  show that the interface
band line--up is not strongly dependent on the interface
morphology and that all the SBH values found are
in qualitative agreement with the experimental value reported
for the atomically clean wurtzite--GaN/Al interface\cite{bermudez}.
 We first discuss the results obtained for the two atomically abrupt
defect-free interfaces (configurations A and B).  Table \ref{schbar}
shows that the $p$-type  SBH is only slightly
reduced (by  0.05 eV, of the same order of magnitude of our numerical
precision) in going from the
N- to the Ga-terminated interface. Similar results
\cite{baldereschini,dandrea2}
were obtained for the GaAs/Al interface, where the $p$-type SBH
was found to be reduced by as much as 0.1 eV in going from the As-
to the Ga-terminated case, in agreement with experiment
\cite{wang,cho}. This trend was attributed \cite{bardi3} to the higher
electron affinity of the anion-terminated compared to the
cation-terminated interface,  which tends to lower the $n$-type SBH.

If we compare the SBH values for the A and C configurations,
we note that the $p$-type SBH is almost unaffected by the
presence of the AlN intralayer (the 0.05 eV difference being of
the order of our numerical precision).  Let us explain our findings.
Since the resulting structural relaxations are negligible, the A and C
structures  differ significantly only for the chemical
influence of the Al atom's presence in the layer closest to the
interface.  We have calculated the difference between the double macroscopic
average of the valence charge density for the A and C structures
(denoted in the following as $\overline{\overline{n}}^{A-C}_{diff}(z)$).
A large difference in the interfacial dipole was not observed due to the
similarity of the electronegativity of the Ga and Al species. We
 evaluated its even and odd contribution relative to the
substituted cation position  taken as zero. The even contribution
of $\overline{\overline{n}}^{A-C}_{diff}(z)$  represents the
chemical difference
between Ga and Al (e.g. the different atomic wave functions) and, being a
quadrupolar periodic charge (the monopole term of the electronic
part being exactly compensated by the nuclear part), that is periodic along
$x$ and $y$  but monolayer thick along $z$, it does not give rise
 to any potential discontinuity
which could affect the SBH. On the other hand, the odd
contribution is expected to represent the charge rearrangement due to
chemical substitution, which should give
rise to charge depletion-accumulation  or,
equivalently, to an extra-dipole that may modify the GaN/Al SBH.
Such a dipole can only arise from the asymmetry around the substituted
cation (due to the presence of the interface), and therefore represents a next
nearest neighbors effect.
In Fig.\ref{fig2}, we  show the even (solid line) and odd (dashed line)
terms of $\overline{\overline{n}}^{A-C}_{diff}(z)$; it is clear that the
odd contribution is very small, relative to
the even one. Therefore, as expected,
$\overline{\overline{n}}^{A-C}_{diff}(z)$ does not
modify the final SBH between GaN and Al.  (Similar
results were obtained from first principles calculations performed on the
GaAs/Al system \cite{dandrea2}).

Let us now consider configuration E.  Since we have just shown that the
presence
of an interface AlN layer (configuration C) in place of the usual GaN layer
(configuration A) does not modify the SBH, in a similar way we expect that the
presence of an interface Ga$_{0.5}$Al$_{0.5}$N layer will not affect
the final SBH as well. This is exactly what happens, as shown by an
essentially identical SBH for configurations A and E
(see Table \ref{schbar}, second row).

We now discuss the transitivity rule \cite{margaritondo} for configuration C.
In Ref. \onlinecite{silvia5} we showed that this simple rule
in the {\em abrupt} system is fulfilled
almost exactly
for  hypothethically ``lattice matched" systems.  In contrast, the
difference in lattice constants inherent to the GaN/AlN/Al system was seen to
 produce deviations of about 0.1 eV, due to differences in structural
interface relaxations \cite{silvia5}.
Let us now observe that configuration C can be seen as a sequence of two
interfaces (as far as a single monolayer may represent most of
electrostatic potential line--up):
a  GaN/AlN heterojunction and an AlN/Al semiconductor/metal
interface.
According to the transitivity rule and recalling the results obtained
for the potential discontinuity in the AlN/Al and GaN/AlN cases
\cite{silvia5},
we should
therefore obtain: $\Phi_{B_p}(GaN/Al)$ =   $\Phi_{B_p}(AlN/Al)$ -
$\Delta\:E_v(GaN/AlN)$ = 1.80 - 0.76 = 1.04 eV, which is in excellent
agreement with
our FLAPW result, $\Phi_{B_p}^C(GaN/Al)$ = 1.07 eV.

Now  consider  the D structure;   note that its $p$-type SBH is very
similar to the value obtained for the other structures.
This can be explained considering that the D and B structures differ
essentially
by the isovalent substitution Ga$\leftrightarrow$Al in the semiconductor
layer closest to the interface. As shown for the A and C structures, the AlN
intralayer does not alter appreciably the SBH; a similar situation can
therefore
be expected upon going from the B to the D structures, as actually found
in Table \ref{schbar}.
In more detail, the SBH variation on going from B to D has opposite sign,
relative to the one seen on going from A to C; this may be related to slightly
different atomic relaxations in the two cases.

Let us now discuss how the SBH can be
altered by displacing the atoms in the metallic side of the interface.
 In order to investigate this subject, we make use of the
 Born effective charges  ($Z^*$) concept,  $Z^*$  being
  the dipole linearly induced by a
 unitary displacement of a single ion in an otherwise perfect crystal
 \cite{kunc}. According to Ref.\onlinecite{alice}, if an atom with  $Z^*$
 is displaced by an amount $u$ from its equilibrium position, the resulting SBH
 change can be expressed as: $\Delta\:V =
 8\:\pi\:e^2\:Z^*\:u/(\varepsilon_{\infty}\:a^2)$, where $\varepsilon_{\infty}$
 is the dielectric constant and $a$ is the cubic lattice constant
 of the medium. Of course, it is expected that in a metal,
 $Z^*$  vanishes due to perfect screening. In order to
 evaluate the $Z^*$  of the metallic Ga and Al atoms, we performed FLAPW
 calculations for some D type structures, obtained by varying the interface
 distance between the Ga$^2_i$ and Al$^1_i$ atoms with respect to the
 equilibrium distance. We found, taking $u$ as large as
$u$ = 0.6 a.u., that the variation of the SBH was less than 0.05 eV. This
confirms, as expected, that the $Z^*$ for the metallic atoms
is nearly vanishing
and that variations of the SBH as a function of the displacements of
 the metallic atoms in the metal side are therefore negligible.

\section{Conclusions}
\label{concl}

First principles calculations have been performed for the GaN/Al
junction, considering various interface
morphologies, namely different atomic terminations for  GaN,
interface atomic exchanges and
Ga$_x$Al$_{1-x}$N intralayers. Our results show that the
Ga$\leftrightarrow$Al atomic exchange has an enthalpy
change of 0.1 eV per atom.

We find that the first few layers of semiconductor are
metallic due to the penetration of MIGS.  For this
reason, the presence of ultra--thin intralayers
 does not result in
a bandgap enhancement near the interface.
In addition,  the effects of the
interface configuration on the electrical properties of the junction
 are found to be  negligible: the   Schottky barrier height
varies by less than 0.1 eV, changing the interface morphology.
In particular, we found that the $p$-type Schottky barrier height
is only slightly
reduced by the presence of {\em (i)} a Ga-termination of the semiconductor
side of the junction, {\em (ii)} a Ga$\leftrightarrow$Al atomic swap at the
immediate interface and {\em (iii)} a Ga$_x$Al$_{1-x}$N ($x$ = 0.5, 1)
intralayer.
Changes in the SBH are explained in terms of the
different character of the
interface bonds in the two Ga- and N- terminated
atomically abrupt cases
and of the Ga$_x$Al$_{1-x}$N ($x$ = 0.5, 1) intralayer
electronic properties.
Our results show good agreement with other theoretical
values for the $p$-type SBH obtained for the
similar GaAs/Al system.

\section{Acknowledgements}
One of the authors (N.N.)  acknowledges  support by the Office of Naval
Research (Contract N00014-96-1002).
This work was supported by a
supercomputing grant at Cineca (Bologna, Italy) through the Istituto Nazionale
di Fisica della Materia (INFM), by the MRSEC program of the National Science
Foundation (DMR-9632472) at the Materials Research Center of
Northwestern University and by a grant of computer time at the
NSF-supported Pittsburgh Supercomputing Center.

\begin{table}
\caption{Structural parameters for the configurations examined
(all values  in atomic units).}
\vspace{5mm}
\begin{tabular}{|c|c|}
Sys. & Interface configuration and interplanar atomic distances \\ \hline
\hline
A & Ga$_i$ $\leftarrow$ \scriptsize 2.12 \normalsize $\rightarrow$
N$_i$ $\leftarrow$ \scriptsize 2.17 \normalsize $\rightarrow$ Al$_i$
$\leftarrow$ \scriptsize 3.67 \normalsize$\rightarrow$ Al \\ \hline
B & N$_i$ $\leftarrow$ \scriptsize 2.23 \normalsize $\rightarrow$
Ga$_i$ $\leftarrow$ \scriptsize 3.04 \normalsize $\rightarrow$ Al$_i$
$\leftarrow$ \scriptsize 3.67 \normalsize$\rightarrow$ Al \\ \hline
C & Ga$_i$ $\leftarrow$ \scriptsize 2.12 \normalsize $\rightarrow$
N$_i^1$ $\leftarrow$ \scriptsize 2.12 \normalsize $\rightarrow$ Al$_i^1$
$\leftarrow$ \scriptsize 2.13 \normalsize$\rightarrow$ N$_i^2$ $\leftarrow$
\scriptsize 2.17 \normalsize $\rightarrow$ Al$_i^2$ $\leftarrow$
\scriptsize 3.67 \normalsize$\rightarrow$ Al\\ \hline
D & Ga$_i$ $\leftarrow$ \scriptsize 2.12 \normalsize $\rightarrow$
N$_i^1$ $\leftarrow$ \scriptsize 2.12 \normalsize $\rightarrow$ Al$_i^1$
$\leftarrow$ \scriptsize 2.12 \normalsize$\rightarrow$ N$_i^2$ $\leftarrow$
\scriptsize 2.12 \normalsize $\rightarrow$ Ga$_i^2$ $\leftarrow$
\scriptsize 3.00 \normalsize$\rightarrow$ Al$_i^2$$\leftarrow$
\scriptsize 3.67 \normalsize$\rightarrow$ Al\\ \hline
E & Ga$_i$ $\leftarrow$ \scriptsize 2.12 \normalsize $\rightarrow$
N$_i^1$ $\leftarrow$ \scriptsize 2.12 \normalsize $\rightarrow$ X$_i$
$\leftarrow$ \scriptsize 2.12 \normalsize$\rightarrow$ N$_i^2$ $\leftarrow$
\scriptsize  2.17 \normalsize $\rightarrow$ Al$_i^2$ $\leftarrow$
\scriptsize 3.67 \normalsize $\rightarrow$ Al
\end{tabular}
\label{struc}
\end{table}

\begin{table}
\caption{$p$-type Schottky barrier heights for the different
$n$ GaN + $m$ Al  interfaces (values in eV).}
\vspace{5mm}
\begin{tabular}{|c|c|c|c|c|c|}
 & A& B& C&D& E \\
$n+m$&GaN/Al N-term. &GaN/Al Ga-term. & AlN intral.&
Ga$\leftrightarrow$Al swap& Ga$_{0.5}$Al$_{0.5}$N intral.  \\ \hline \hline
15+7 & 1.12 & 1.06 & 1.07 & 1.11 & -\\
11+5 & 1.30  & - & - & - & 1.30\\
\end{tabular}
\label{schbar}
\end{table}

\begin{figure}
\caption{Different interface geometries considered.
Small squares regions indicate the GaN substrate;
large square regions represent the Al overlayer. Abrupt
GaN/Al N-terminated interface - system A (panel (a)); abrupt
GaN/Al Ga-terminated interface  - system B (panel (b)); AlN intralayer
 - system C (panel (c)); system D showing the Ga $\leftrightarrow$ Al
atomic swap (panel (d); GaAlN alloy intralayer - system E (panel (e).}
\label{figs}
\end{figure}

\begin{figure}
\caption{Valence charge density projected on planes
cutting   different interface bonds:
(a) bulk Al atoms in configuration B;  (b) Ga$_i$-Al$_i$ atoms
in configuration B; (c)
Ga$_i^2$-Al$_i^2$ atoms in configuration D. Contour lines are spaced by
0.01 electrons/cell.}
\label{chplot}
\end{figure}

\begin{figure}
\caption{Even (solid line) and odd (dashed line) contributions to
$n_{diff}^{A-C}(z)$, defined as the difference between the double macroscopic
average of the valence charge density for the A and C configurations.
Units in electrons/basis-area.}
\label{fig2}
\end{figure}


\begin{thebibliography}{100}
\bibitem{notaform}
Recall that the energy required to form, for
example, a  (GaN)$_n$/(Al)$_m$
superlattice (SL), with $n$ GaN
layers and $m$ Al layers starting from the binary constituents can be
expressed as
$\Delta\:H=|E_{tot}^{SL}| - n \cdot |E_{tot}^{GaN}|- m \cdot |E_{tot}^{Al}|$
where the total energies of the binaries are obtained at the same strain
conditions as in the superlattice.
\end{thebibliography}
\end{document}